\documentclass[review]{elsarticle}

\usepackage{graphicx,amssymb}
\usepackage{epstopdf}
\newcommand{\webb}{{\it Webb}}
\newcommand{\all}{{\it All}}
\journal{Physics Letters B}

\begin{document}
\begin{frontmatter}

\title{Constraining spatial variations of the fine-structure constant in symmetron models}
\author[inst1,inst2,inst3]{A. M. M. Pinho}\ead{am.pinho@thphys.uni-heidelberg.de}
\author[inst4]{M. Martinelli}\ead{martinelli@lorentz.leidenuniv.nl}
\author[inst1,inst5]{C. J. A. P. Martins\corref{cor1}}\ead{Carlos.Martins@astro.up.pt}
\address[inst1]{Centro de Astrof\'{\i}sica, Universidade do Porto, Rua das Estrelas, 4150-762 Porto, Portugal}
\address[inst2]{Faculdade de Ci\^encias, Universidade do Porto, Rua do Campo Alegre 687, 4169-007 Porto, Portugal}
\address[inst3]{Institut f\"ur Theoretische Physik, Ruprecht-Karls-Universit\"at Heidelberg, Philosophenweg 16, 69120 Heidelberg, Germany}
\address[inst4]{Institute Lorentz, Leiden University, PO Box 9506, Leiden 2300 RA, The Netherlands}
\address[inst5]{Instituto de Astrof\'{\i}sica e Ci\^encias do Espa\c co, CAUP, Rua das Estrelas, 4150-762 Porto, Portugal}
\cortext[cor1]{Corresponding author}

\begin{abstract}
We introduce a methodology to test models with spatial variations of the fine-structure constant $\alpha$, based on the calculation of the angular power spectrum of these measurements. This methodology enables comparisons of observations and theoretical models through their predictions on the statistics of the $\alpha$ variation. Here we apply it to the case of symmetron models. We find no indications of deviations from the standard behavior, with current data providing an upper limit to the strength of the symmetron coupling to gravity ($\log{\beta^2}<-0.9$) when this is the only free parameter, and not able to constrain the model when also the symmetry breaking scale factor $a_{SSB}$ is free to vary.
\end{abstract}


\end{frontmatter}

\section{Introduction}\label{sec:intro}

Astrophysical tests of the stability of dimensionless fundamental couplings such as the fine-structure constant $\alpha$ are a powerful probe of cosmology as well as of fundamental physics \cite{Uzan,GRG}. The analysis of a dataset of 293 archival data measurements from the Keck and VLT telescopes by Webb {\it et al.} provided an indication of spatial variations with an amplitude of a few parts per million, with a statistical significance of $4-\sigma$ \cite{webbdipole}. Even though there are concerns about possible systematic effects in this dataset \cite{Whitmore} and the statistical significance itself decreases when this dataset is analyzed jointly with more recent data \cite{Dipoles}, it is important to consider the theoretical implications of such results, also bearing in mind that forthcoming astrophysical facilities will enable much more precise tests in the near future.

At a phenomenological level it is common to fit the astrophysical measurements with a simple dipole, with or without an additional dependence on redshift or look-back time \cite{webbdipole,Dipoles}. On the other hand, from a theoretical point of view simplistic dipole models would require strong fine-tuning to explain such a behavior, and a physically motivated approach would rely on environmental dependencies \cite{olivepospelov}. This therefore calls for more robust methodologies which enable accurate comparisons between models and observations. Early work along these lines was done by Murphy {\it et al.}, who calculated the two-point correlation function of the Keck subsample of the aforementioned archival data, finding it to be consistent with zero \cite{MWF}. 
In this paper we move from the two point angular correlation function to the calculation of the angular power spectrum of these measurements.
The aim of adopting this approach is to be able to compress the data information in such a way to allow for comparison with the predictions of theoretical models.
As a proof of concept, in this paper we apply this method to the case of the symmetron model, for which the environmental dependence of $\alpha$ has been previously studied using N-body simulations \cite{marvin}.

In Section \ref{sec:symm} we present a concise overview of the symmetron model.
Section \ref{sec:data} presents the methodology used to compress the $\alpha$ measurements
into angular power spectra. In section \ref{sec:ana} we calculate the theoretical power
spectrum for the symmetron model and present our analysis methodology,
leading to the results discussed in Section \ref{sec:res}. Finally, in Section \ref{sec:conc} we
summarize our results and the outlook for this methodology.

\section{Symmetron model}\label{sec:symm}

The symmetron model is a scalar-tensor modification of gravity, introduced in order to achieve an additional
long range scalar force while still satisfying local gravity constraints thanks to the environment density 
dependence of its coupling to matter. This modification of gravity is described by the action \cite{Hinterbichler}
\begin{equation}
S = \int dx^4 \sqrt{-g} \Big[ \frac{R}{2} M^2_{pl} - \frac{1}{2} (\partial \phi)^2 - V (\phi) \Big] + S_m( \Psi_m;g_{\mu\nu}A^2(\phi))
\end{equation}
where $g =\det(g_{\mu\nu}), M_{pl}=1/\sqrt{8 \pi G}$ and $S_m$ is the matter-action. The conformal coupling between the scalar field and the matter fields $\Psi_m$ expressed by $\tilde{g}_{\mu\nu} = g_{\mu\nu} A^2(\phi) $, is assumed to be the simplest one consistent with the potential symmetry,
\begin{equation}
A(\phi) = 1 + \frac{1}{2} \Big( \frac{\phi}{M} \Big)^2\,,
\end{equation}
with $M$ and $\mu$ arbitrary mass scales. This coupling leads to a fifth force, which in the non-relativistic limit is given by
\begin{equation}
\overrightarrow{F}_\phi \equiv \frac{d A(\phi)}{d\phi} \overrightarrow{\nabla} \phi = \frac{\phi \overrightarrow{\nabla} \phi}{M^2}.
\end{equation}
The potential is chosen to be of the symmetry breaking form 
\begin{equation}
V(\phi) = -\frac{1}{2} \mu^2 \phi^2 + \frac{1}{4} \lambda \phi^4,
\end{equation}
 The dynamics of the scalar field $\phi$ is determined by an effective potential which in the non-relativistic limit (relevant for the astrophysical measurements) has the form
\begin{equation}
V_{eff} (\phi) = V(\phi) + A(\phi) \rho_m = \frac{1}{2} \Big( \frac{\rho_m}{\mu^2 M^2} - 1 \Big) \mu^2 \phi^2 + \frac{1}{4} \lambda \phi^4\,;
\end{equation}
this means that in the early Universe or, in general, when the matter density is high, the effective potential has a minimum $\phi = 0$ where the field will reside. As the Universe expands, the matter density dilutes until it reaches a critical density $\rho_{SSB} = \mu^2 M^2$ for which the symmetry breaks and the field moves to one of the two new minima $\phi = \pm \phi_0 = \pm \mu / \sqrt{\lambda}$. \\

The fifth-force between two test particles residing in a region of space where the field has the value $\phi=\phi_{local}$ can be calculated to be \cite{Hinterbichler}
\begin{equation}
\frac{F_\phi}{F_{gravity}} = 2 \beta^2 \Big( \frac{\phi_{local}}{\phi_0} \Big)^2\sim2\beta^2\left(1-\frac{\rho}{\mu^2M^2}\right)\,,
\end{equation}
for separations of the Compton wavelength $\lambda_{local} = 1/\sqrt{V_{eff,\phi \phi} (\phi_{local})}$, where the coupling strength to gravity is given by
\begin{equation}
 \beta = \frac{\phi_0 M_{pl}}{M^2}
\end{equation}
For larger separations or in the cosmological background before symmetry breaking, $\phi_{local} \approx 0$ and the force is suppressed. After symmetry breaking, the field moves towards $\phi = \pm \phi_0$ and the force is comparable to gravity for $\beta = \mathcal{O} (1)$. Non-linear effects in the field-equation ensure that the force is effectively screened in high density regions. The symmetry breaks at the scale factor $a_{SSB} =(\rho_{m,0}/\rho_{SSB})$ and the range of the fifth-force when the symmetry is broken is given by $\lambda_{\phi 0} = 1/(\sqrt{2} \mu)$, where local gravity constrains satisfy $\lambda_{\phi 0} \lesssim  1$ Mpc/h for symmetry breaking close to today, i.e. $a_{SSB} \approx 1$ \cite{symcosmo}.

\begin{figure}[!h]
\begin{center}
\includegraphics[width=\columnwidth]{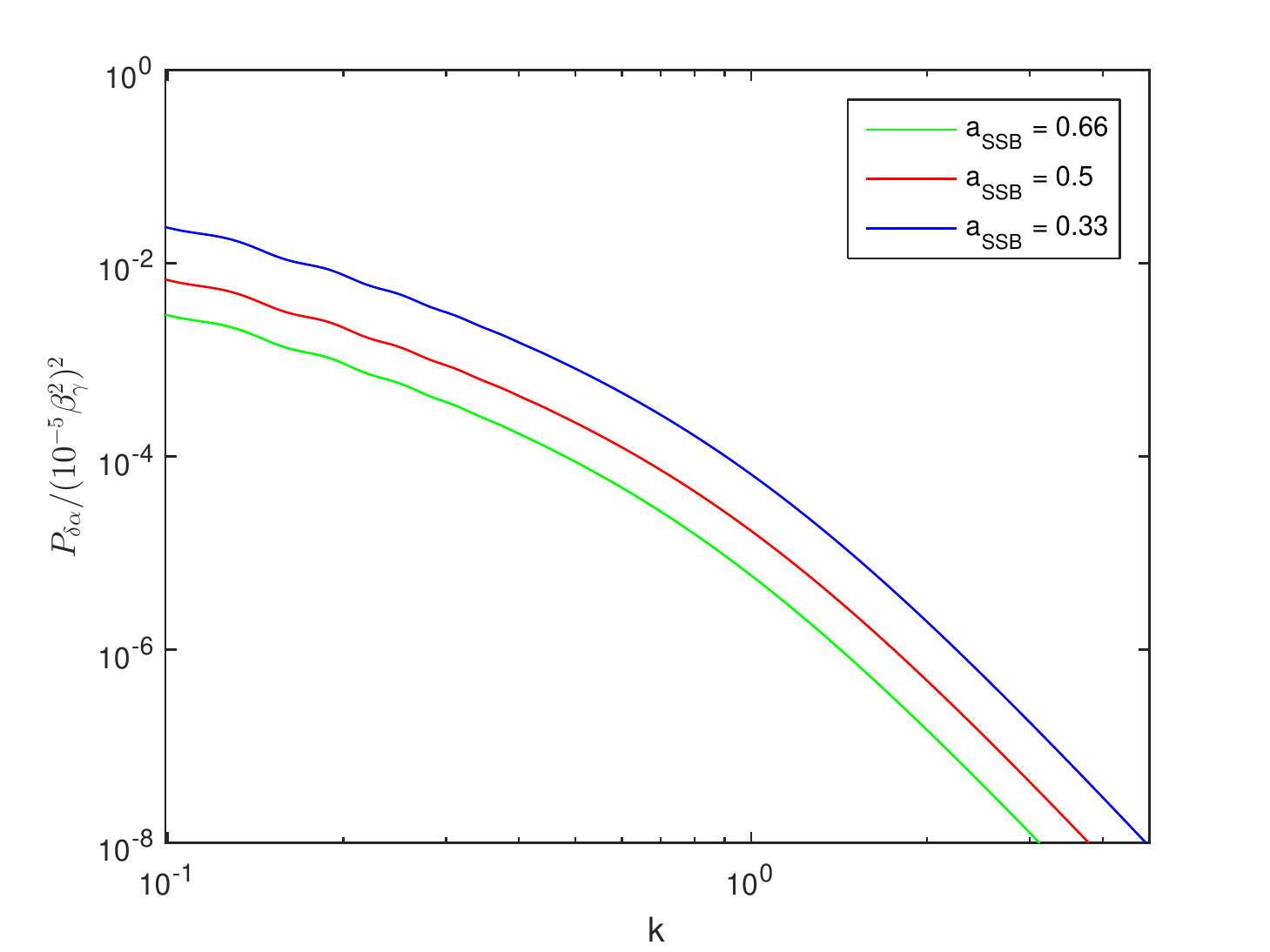}
\caption{Theoretical power spectrum $P_{\delta\alpha}(k,a)$ given by eq. \ref{eq.PS} as a function of the wavenumber $k$ for $a = 1, \beta = 1, \lambda_{\phi 0} = 1$ Mpc/h and different symmetry breaking scale factors $a_{SSB} = [0.33, 0.5, 0.66]$. Note that strictly speaking eq. \ref{eq.PS} only applies in the linear regime, so the behaviour beyond this should be taken with care. Following \protect\cite{marvin} a normalization factor $x = 0.06 (0.5/a_{SSB}$) was used.}
\label{plt.thPS}
\end{center}
\end{figure}

Since the symmetron scalar field is a dynamical degree of freedom, one naturally expects it to couple to the other degrees of freedom in the Lagrangian, unless a new symmetry is postulated to suppress these couplings. In particular, we can assume that it couples with the electromagnetic sector of the theory \cite{marvin}
\begin{equation}
S_{EM} = - \int dx^4 \sqrt{g} B_F(\phi) \frac{1}{4} F_{\mu\nu}^2\,,
\end{equation}
where $B_F$ is the gauge kinetic function which leads to $\alpha = \alpha_0 B_F^{-1} (\phi)$. With the same choice of quadratic coupling $B_F^{-1} (\phi) = 1 + \frac{1}{2} \beta_\gamma^2 \Big( \frac{\phi}{M} \Big)^2$ one gets the following variation of the fine structure constant
\begin{equation}
\delta_\alpha\equiv \frac{\Delta \alpha}{\alpha}= \frac{\alpha(\phi)-\alpha_0}{\alpha_0} = B_F^{-1}(\phi) -1 =  \frac{1}{2} \Big( \frac{\beta_\gamma \phi}{M} \Big)^2.
\end{equation}
Considering perturbations of the scalar field in Fourier space, the power spectrum for variations of $\alpha$ in the linear regime can be connected to the matter power spectrum $P_m (k,a)$ as follows \cite{marvin}
\begin{equation}
P_{\delta_\alpha}(k,a) = \Bigg[ \frac{3 \Omega_m H_0^2 \beta_\gamma^2 \beta^2}{a (k^2 + a^2 m_\phi^2)} \Bigg( \frac{\overline{\phi}}{\phi_0} \Bigg)^2 \Bigg]^2 P_m (k,a).
    \label{eq.PS}
\end{equation}
where $\Omega_m$ and $H_0$ are the present-day matter density and Hubble parameter, $\beta_\gamma$ is the scalar-photon coupling relative to the scalar-matter coupling, $k$ is the co-moving wavenumber, $m_\phi^2 = V_{eff,\phi \phi}(\bar{\phi})$ is the scalar mass in the cosmological background, and $(\overline{\phi} / \phi_0)$ is the background scalar field value. For $a \geq a_{SSB}$ we can write
\begin{equation}\label{eq:Pdalpha}
\Bigg( \frac{\overline{\phi(a)}}{\phi_0} \Bigg)^2 = \Bigg( 1 - \Big( \frac{a_{SSB}}{a} \Big)^3 \Bigg), \qquad 
	m_\phi^2 (a) =\frac{1}{\lambda_{\phi 0}^2} = \Bigg( 1 - \Big( \frac{a_{SSB}}{a} \Big)^3 \Bigg).
\end{equation}
$P_{\delta_\alpha}$ is plotted in figure \ref{plt.thPS}; it is also useful to write it as
\begin{equation}
P_{\delta_\alpha}(k,a) = \Bigg[ \frac{0.33 \Omega_m 10^{-6} \beta_\gamma^2 \beta^2}{a ((k/m_\phi)^2 + a^2)} \Bigg( \frac{\lambda_{\phi 0}}{Mpc / h} \Bigg)^2 \Bigg]^2 P_m (k,a).
    \label{eq.lPS}
\end{equation}

\section{Observational data}\label{sec:data}

Currently available astrophysical measurements of $\alpha$ come from high-resolution spectroscopy of absorption clouds along the line of sight of bright quasars. In addition to the 293 archival measurements of Webb {\it et al.} \cite{webbdipole} there are 20 more recent dedicated measurements discussed in \cite{Dedicated}, making a total of 313 measurements. From now on we refer to the former as \webb\ and as \all\ to the combination of these with the latter. For each of them, apart from the measurement of the relative variation of $\alpha$ itself, the sky coordinates and redshift of the absorber (spanning the range $0.2<z<4.2$) are known with negligible uncertainty. We can compress this information in an angular power spectrum $C_\ell$, to be compared with statistical predictions coming from theoretical models.

In order to do this, we obtain the two-point correlation function $c(\vartheta)$ from the $\delta_\alpha(\theta,\phi)$ measurements \cite{MWF}
\begin{equation}\label{eq:nusscorr}
        c(\vartheta) =  \frac{1}{\bar{n}^2 f_{sky}} < \delta_\alpha(\theta,\phi) \delta_\alpha(\theta',\phi')>,
\end{equation}
where the brackets $< . >$ correspond to the average taken over all possible orthodromic separations $\vartheta$. Since the measurements of $\delta_\alpha$ are sparse on the sky (effectively point sources), the discreteness of the data has been taken into account following the procedure of \cite{nusser}: $4 \pi f_{sky}$ steradians is the assumed  coverage of the sky of the dataset and $\bar{n} = N/(4 \pi f_{sky})$ is the corresponding mean number density over the observed part of the sky (with N as the number of sources).The data we consider here are effectively sparse point sources spread over the whole sky, therefore we take $f_{sky} = 1$.
This assumption provides a conservative estimate of the measurements density $\bar{n}$ and therefore of the power spectrum estimation, despite also affecting cosmic variance, decreasing its impact on the estimator noise. Future more complete datasets will allow to deal properly with these aspects, exploiting also techniques commonly used for other cosmological observables, such as CMB or galaxy surveys, and therefore to obtain a more precise estimate of the error contributions.\\
In this work we are also neglecting the redshift information of the $\delta_\alpha$ measurements; in practice we are assuming that $\delta_\alpha$ has no redshift dependence and all the deviations from the standard value are brought by spatial variations. This approach is acceptable with the current state of the data, but it will be crucial to include redshift information when the data will reach a sensitivity allowing for a tomographic reconstruction of $\delta_\alpha$. Moreover, including the possibility of $\delta_\alpha(z)$ will be necessary to test theoretical models which also predict a time evolution for the fine structure constant.

We can in principle perform a Legendre transform of the angular correlation in order to obtain the angular power spectrum $C_\ell$ as\cite{loverde}
\begin{equation}
C_\ell = \int c(\vartheta) P_\ell (cos \vartheta) d \Omega.
\end{equation}
where $P_\ell ( cos \vartheta)$  is the Legendre polynomial and $\Omega$ the solid angle. In practice, we compute the power spectrum estimator $\hat{C}_\ell$ as
\begin{equation}
\hat{C}_\ell = 2 \pi \sum_\vartheta c(\vartheta) P_\ell (cos \vartheta) sin \vartheta \Delta \vartheta
\label{eq.cl}
\end{equation} 
with $\Delta \vartheta$ being the difference between consecutive values of the angular separation $\vartheta$. The expected error of the power spectrum estimator can be obtained from \cite{nusser}
\begin{equation}
\Sigma^2 = \frac{2}{(2l+1) f_{sky}} \Big( \frac{\sigma_f^2}{\bar{n}} + \hat{C}_\ell \Big)^2
\label{eq.Sigma}
\end{equation}
which includes both contributions of the shot noise $\Sigma_{SN}$ and cosmic variance $\Sigma_{CV}$ that can be expressed as
\begin{equation}
\Sigma_{SN} = \sqrt{\frac{2}{(2l+1) f_{sky}}} \frac{\sigma_f^2}{\bar{n}}, \qquad
\Sigma_{CV} = \sqrt{\frac{2}{(2l+1) f_{sky}}} \hat{C}_\ell.
\label{eq.SNCV}
\end{equation}
$\sigma_f$ is obtained from the measurements' errors\footnote{if both systematic and statistical errors are known, we use the combined error obtained by adding them in quadrature} $\sigma_j$ weighting each measurement with a factor $w_i^2$ given by
\begin{equation}
	w_i^2 = \frac{N \sigma_i^{-2}}{\sum_j \sigma_j^{-2}}
\end{equation}
which yields the aforementioned quantity $\sigma_f$ as 
\begin{equation}
\sigma_f^2 = \frac{N}{\sum_j \sigma_j^{-2}}.
\label{eq.sigfw}
\end{equation}

Often there are several measurements of $\delta_\alpha$ at different redshifts along the same line of sight as the light from the quasar can go through more than one absorption cloud until it reaches Earth. To avoid null angular separations in the computation of Eq. (\ref{eq:nusscorr}), we choose to use the weighted mean measurement for measurements in the same line of sight. Our full dataset includes measurements from 156 independent lines of sight. Before computing the correlation function, the dataset is analyzed and replaced by new values of weighted redshift, $z_w$, weighted $\delta_{\alpha,w}$ and its corresponding weighted error, $\sigma_w$ described by
\begin{equation}
	z_w = \frac{\sum_i w_i \times z_i}{\sum_i w_i}, \qquad 
    \frac{\Delta \alpha}{\alpha} \bigg \vert_w = \frac{\sum_i w_i \times \frac{\Delta \alpha}{\alpha}_i}{\sum_i w_i}, \qquad
    \sigma_w^2 = \frac{1}{\sum_i w_i}
\end{equation}
where $w$ is the weight given by $w_i = 1/\sigma_i^2$ and the index $i$ runs over the measurements on each line of sight.

\begin{figure}[h!]
 \begin{center}
  \begin{tabular}{cc}
   \includegraphics[width=0.5\columnwidth]{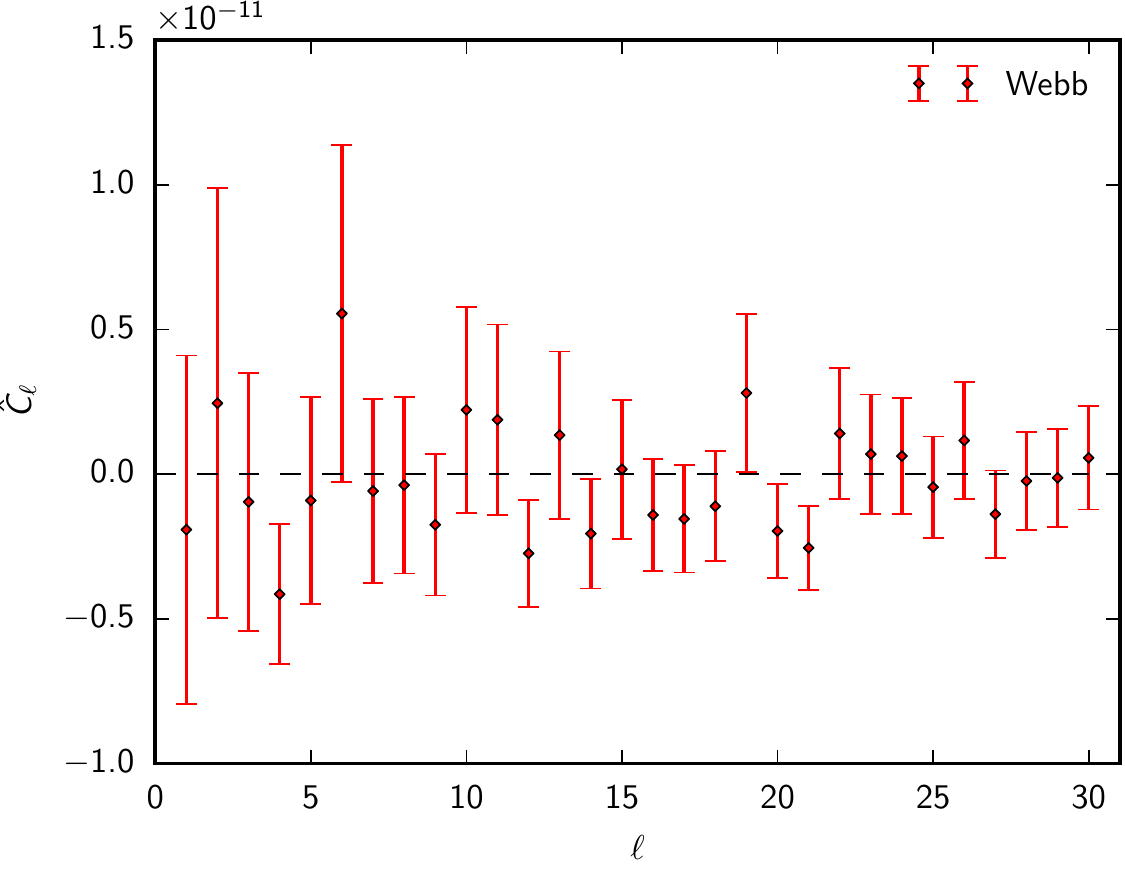} &
   \includegraphics[width=0.5\columnwidth]{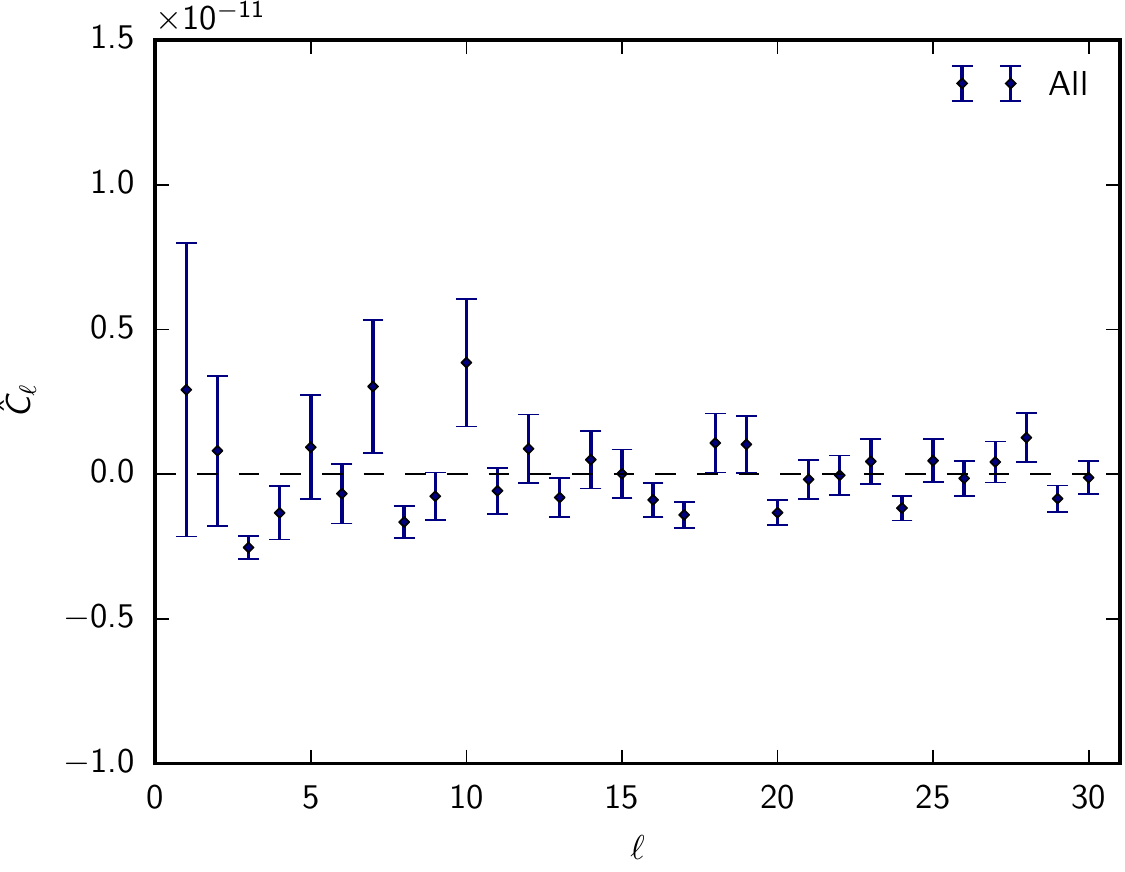} \\
  \end{tabular}
  \caption{Angular power spectrum estimation $\hat{C}_\ell$ as a function of the multipole $\ell$ with its expected error $\Sigma$ for the \webb\ dataset , and for the \all\ combination.}
  \label{plt.clforall-fig3}
 \end{center}

\end{figure}

\begin{figure}[h!]
 \begin{center}
  \begin{tabular}{cc}
   \includegraphics[width=0.5\columnwidth]{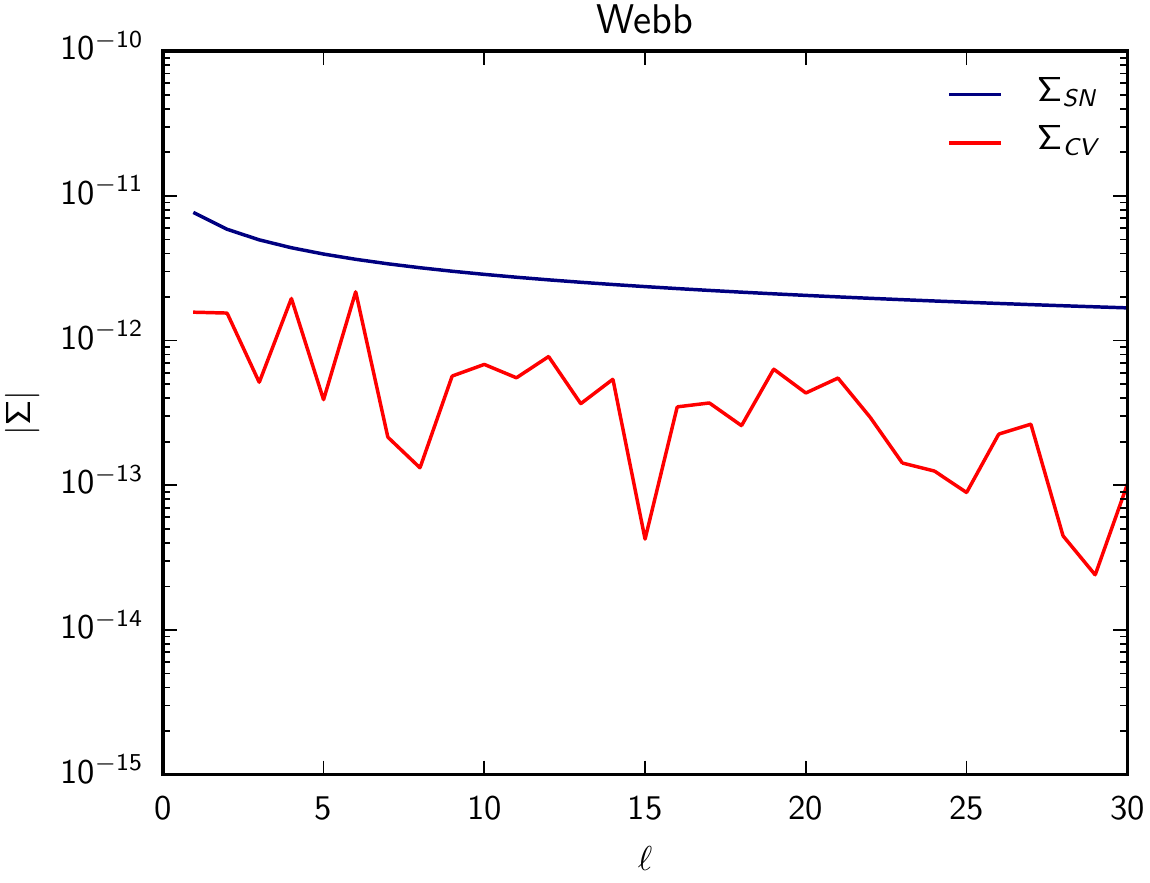} &
   \includegraphics[width=0.5\columnwidth]{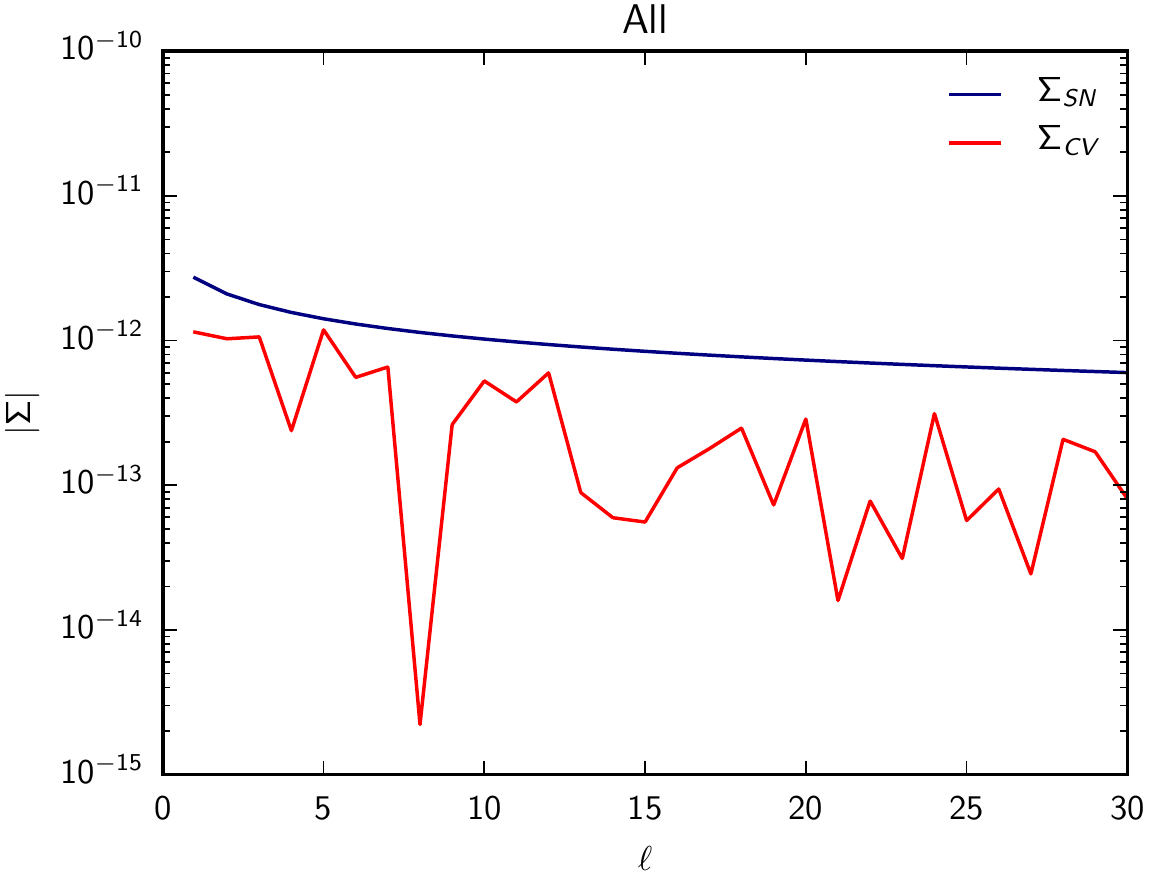} \\
  \end{tabular}
  \caption{Contributions to the error $\Sigma$ coming from shot noise $\Sigma_{SN}$ (blue lines) and cosmic variance $\Sigma_{CV}$ (red lines). The left panel refers to the \webb\ dataset while the right one shows the case of the \all\ combination.}
  \label{plt.clforall-fig9}
 \end{center}

\end{figure}

Figure \ref{plt.clforall-fig3} shows the angular power spectra $\hat{C}_\ell$ obtained with the procedure described above, considering the \webb\  dataset (left panel)
and the \all\ combination (right panel). We notice how the inclusion of the new data, although limited in the number of sources, leads to an improvement of the measured power spectrum, thanks to the increased sensitivity.\\
Figure \ref{plt.clforall-fig9} shows instead the global error on the measurements and the contributions coming from shot noise and cosmic variance, where we can notice how the former dominates over the latter even at the large scales considered.\\

\section{Theoretical predictions and data analysis}\label{sec:ana}

\begin{figure}
\begin{center}
\begin{tabular}{cc}
\includegraphics[width=0.5\columnwidth]{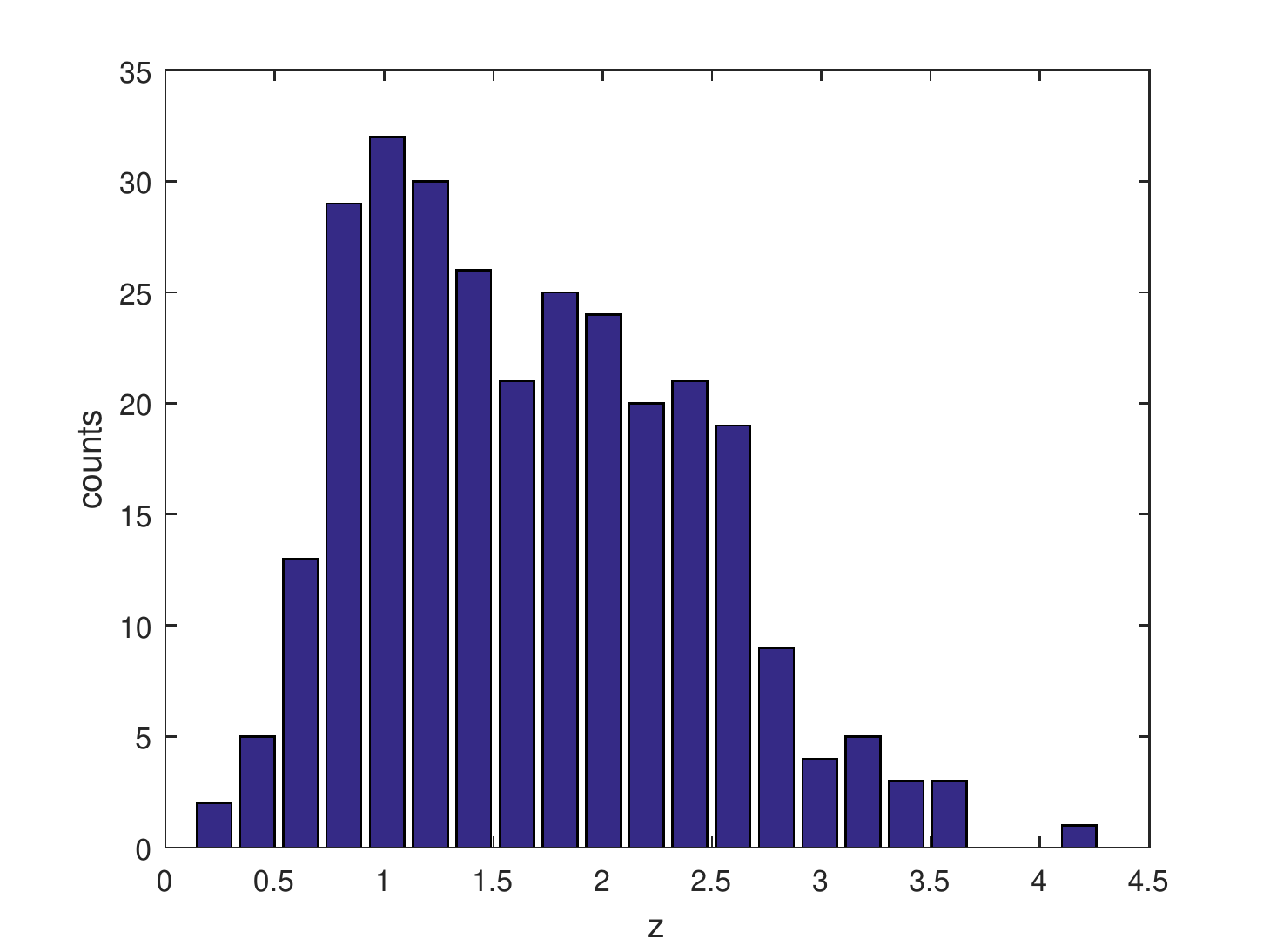} &
\includegraphics[width=0.5\columnwidth]{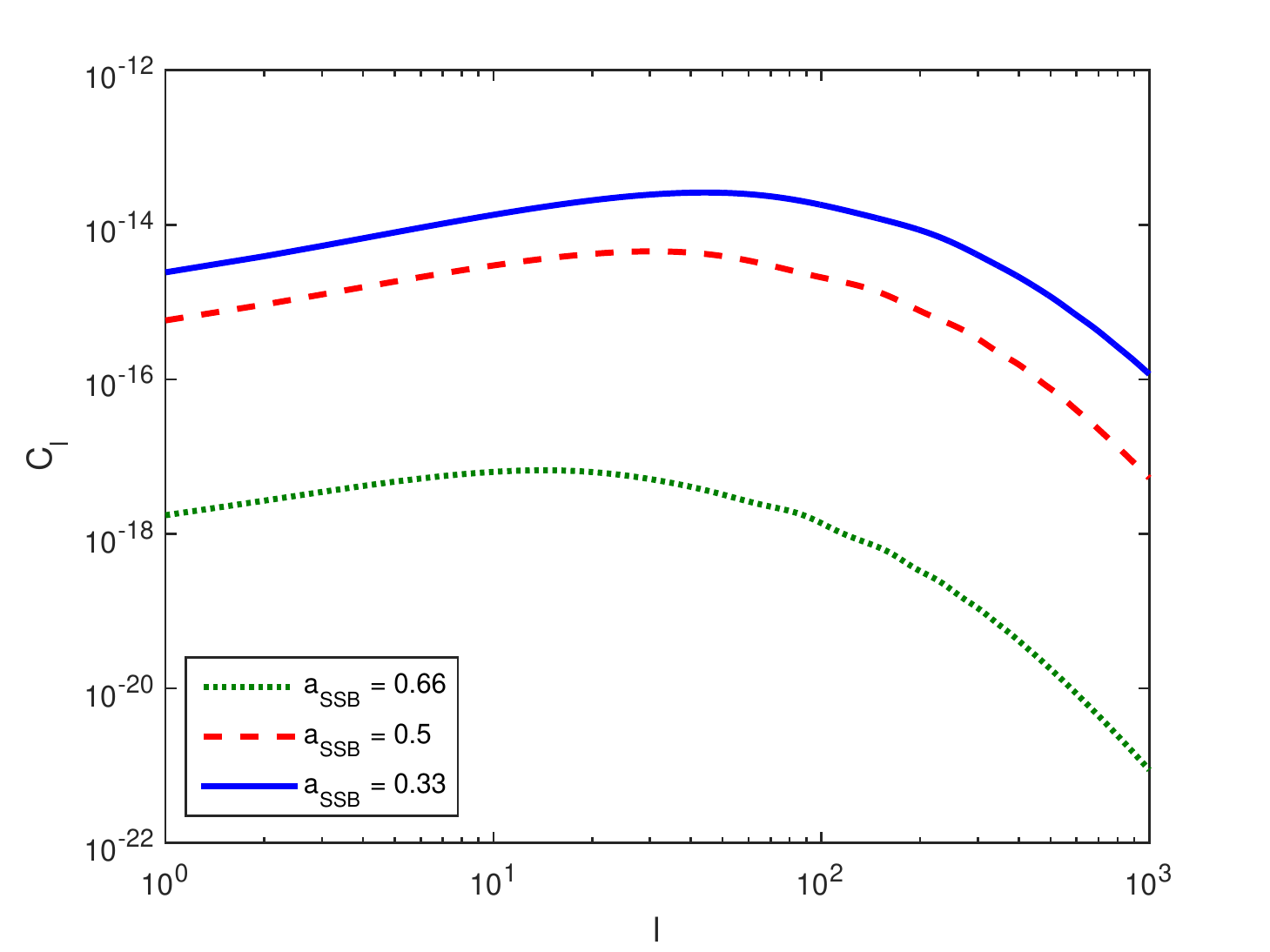}
\end{tabular}
\caption{Left panel: Source distribution function in redshift space for the archival dataset of \protect\cite{webbdipole}. Right panel: Theoretical power spectrum $C_\ell$ for the symmetron model for different values of the scale factor for the symmetry breaking $a_{SSB}$.}
\label{plt.pWebb}
\end{center}
\end{figure}

We can now compare the observational power spectra with the predictions made by the symmetron model. This entails expressing Eq.~(\ref{eq.PS}) in the form of an angular power spectrum. Generically this can be written as 2D projection of the 3D density field which in this case is the linear power spectrum $P(k,z)$. \\
In this paper we exploit the Limber approximation, which simplifies the calculations by avoiding integrations
of Bessel functions. We warn the reader that this approximation can significantly impact the calculation of the power spectra \cite{loverde}, especially at the angular scales considered here; however, since the sensitivity of the currently available data is far from allowing precision reconstructions of $P_{\delta_\alpha}$, the 
use of more refined methods is outside the aim of this paper, but this should be taken into account when more precise measurements will be available.
In this approximation, we can compute the angular power spectrum as \cite{loverde}
\begin{equation}\label{eq:thcl}
C_\ell \approx \int dz W^2(z) \frac{H(z)}{d_A^2(z)} P_{\delta\alpha} \Bigg( k = \frac{l+1/2}{r}; z \Bigg)
\end{equation}
where $W(z)$ is the normalized source distribution function in redshift space, 
$H(z)$ is the Hubble parameter function, $d_A(z)$ is the angular diameter function and $P(k,z)$ is the linear power spectrum previously obtained in Eq.~(\ref{eq.PS}), with $k = \frac{l+1/2}{r}$, and $r$ is the comoving distance. We reconstruct the source distribution function with a 20 bins histogram from each dataset considered. The example of the archival dataset source distribution can be found in figure \ref{plt.pWebb}.

In order to compare these theoretical spectra with the observational data, we use the publicly available code \texttt{COSMOMC} \cite{cosmomc}, modified in such
a way to compute from a set of parameters the theoretical power spectrum of Eq.~(\ref{eq:thcl}), where $P_{\delta\alpha}$ is given by Eq.~(\ref{eq:Pdalpha}).

We consider here as free parameters the scale factor when the symmetry breaks ($a_{SSB}$) and the product of the coupling $\beta_\gamma\beta$ (from now on simply named $\beta$ for simplicity), while we fix the range of the fifth force when the symmetry is broken to $\lambda_{\phi 0} = 1$ Mpc/h: this choice saturates the local gravity constraint of \cite{symcosmo} and was also used in the N-body simulations of \cite{marvin}. In principle, also the standard cosmological parameters should be included in the analysis as they affect the calculation of the power spectra through Eqs. 
(\ref{eq.PS},\ref{eq:thcl}). However, current $\alpha$ datasets would not be able to simultaneously constrain the standard and symmetron parameters, therefore additional observables such as CMB or supernovae should be included, provided the impact of a spatial variation of $\alpha$ is included also in the analysis of these. We decide to leave
these considerations for a future, more detailed, paper on the topic and rather take for the standard cosmological parameters the marginalized value from Planck 2015 \cite{planck}, focusing only on constraints of the symmetron parameters.

\section{Results}\label{sec:res}

\begin{figure}
\begin{center}
\begin{tabular}{cc}
\multicolumn{2}{c}{
\includegraphics[width=.5\columnwidth]{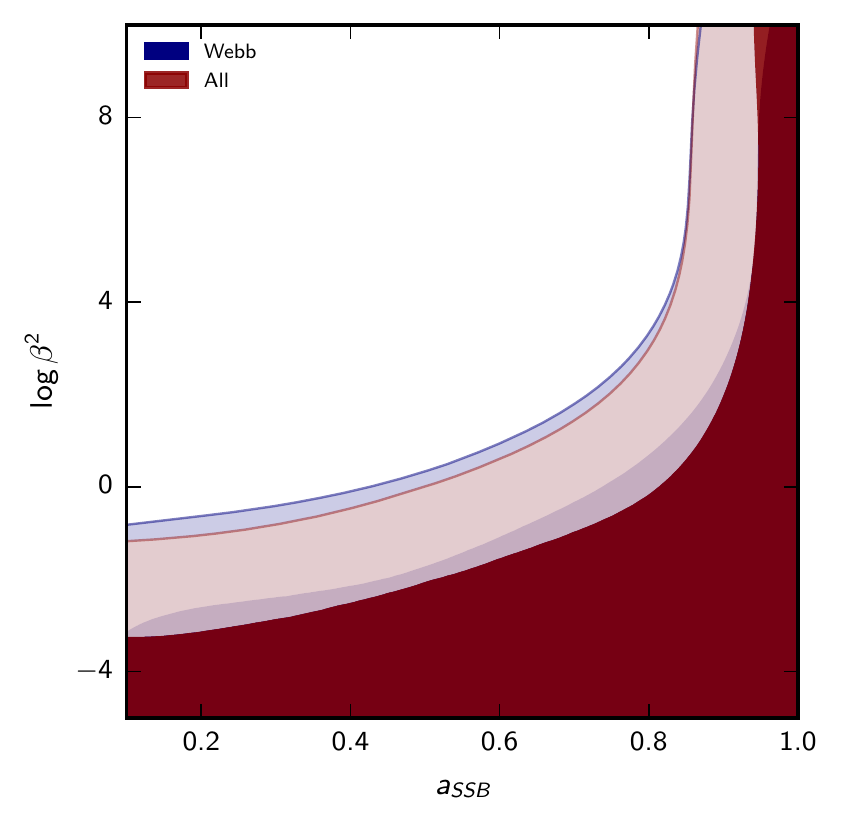}}\\
\includegraphics[width=.5\columnwidth]{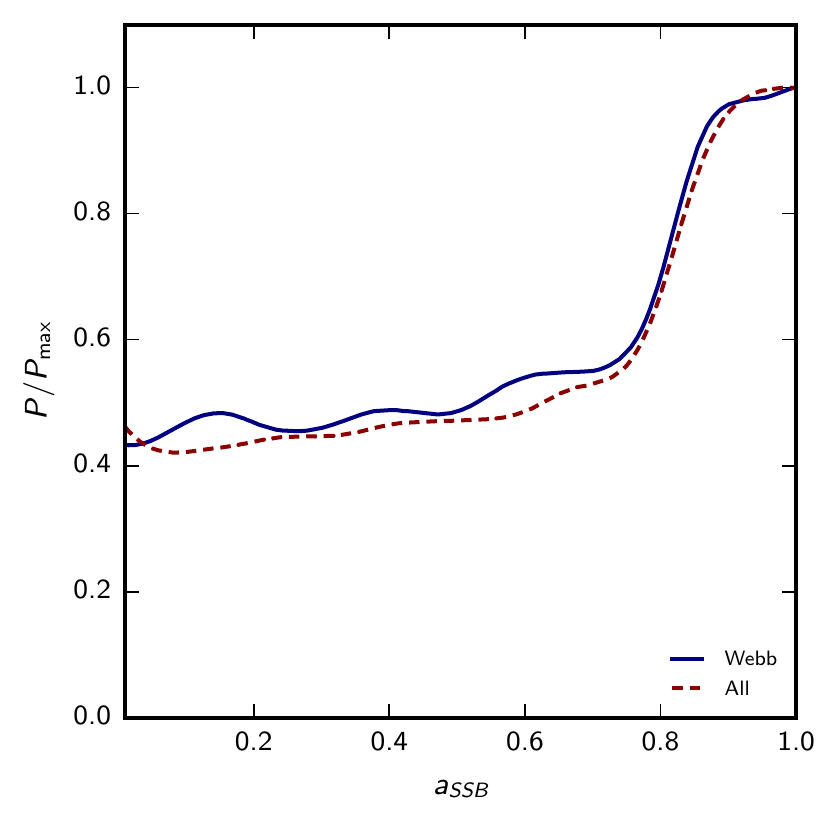}&
\includegraphics[width=.5\columnwidth]{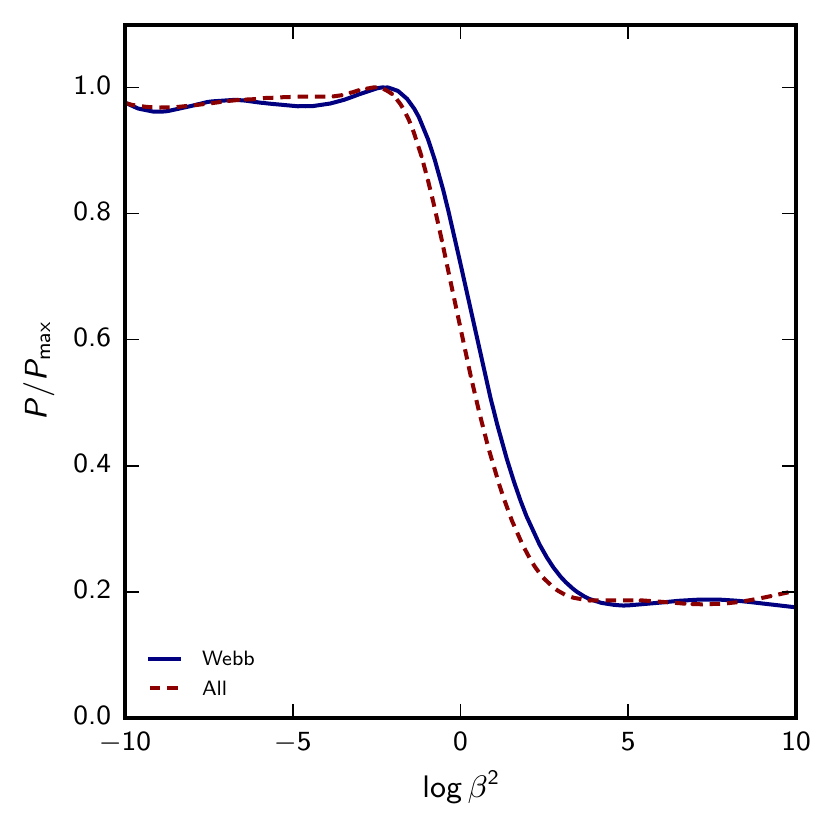}\\
\end{tabular}
\caption{{\it Top panel:} $1$ and $2-\sigma$ confidence regions in the $a_{SSB}-\log{\beta^2}$ plane obtained using the \webb\ (blue contours) and the \all\ (red contours) datasets. {\it Bottom panels:} posterior distributions for $a_{SSB}$ (left panel) and $\log{\beta^2}$ (right panel), where blue solid lines refer to the results obtained using the \webb\ dataset and red dashed ones refer to the \all\ combination.}
\label{plt.webb2p}
\end{center}
\end{figure}

In this section we highlight the results obtained applying the analysis described above. In our first analysis, we allow both $a_{SSB}$ and $\log{\beta^2}$ to vary assuming flat prior distributions\footnote{we sample $\log{\beta^2}$ instead of $\beta^2$ in order to better sample the low coupling limit}, while fixing $\lambda_{\phi 0}=1$ Mpc/h. Figure \ref{plt.webb2p} shows the posterior distribution for each of the two free parameter using different datasets.

Overall we find that the \webb\ dataset is not able to constrain the two parameters simultaneously and no deviations from a vanishing $\alpha$ variation are found. As the recent dedicated measurements are consistent with a non-varying $\alpha$, when combined with the larger archival dataset, they lead again to an agreement with a vanishing $\beta^2$, but they are still not able to put bounds on the parameters.\\

\begin{figure}
\begin{center}
\begin{tabular}{cc}
\multicolumn{2}{c}{
\includegraphics[width=0.5\columnwidth]{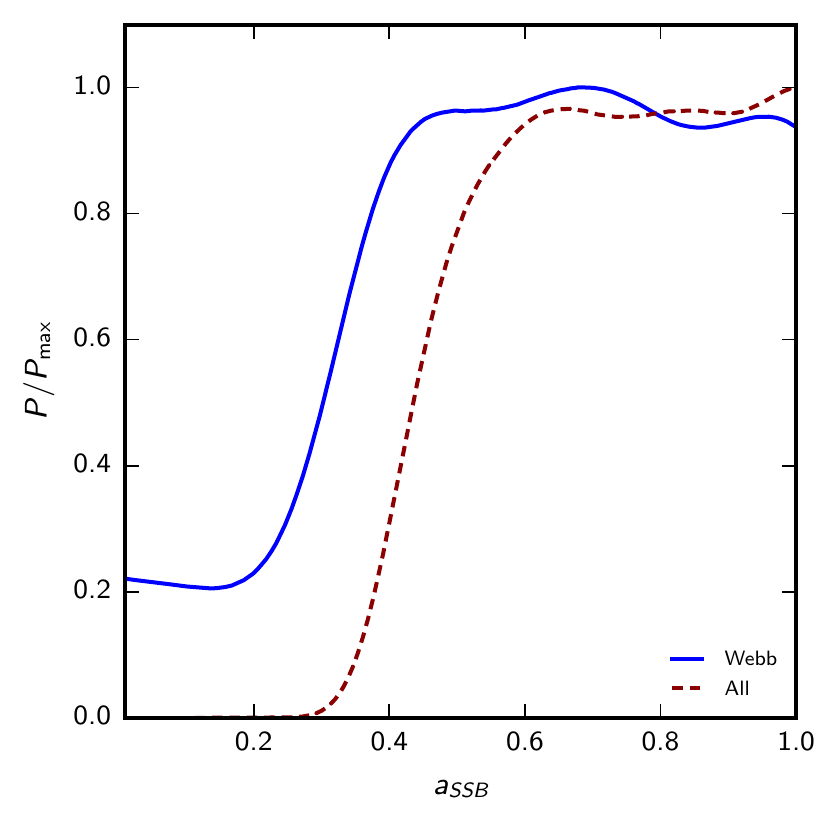}}\\
\includegraphics[width=0.5\columnwidth]{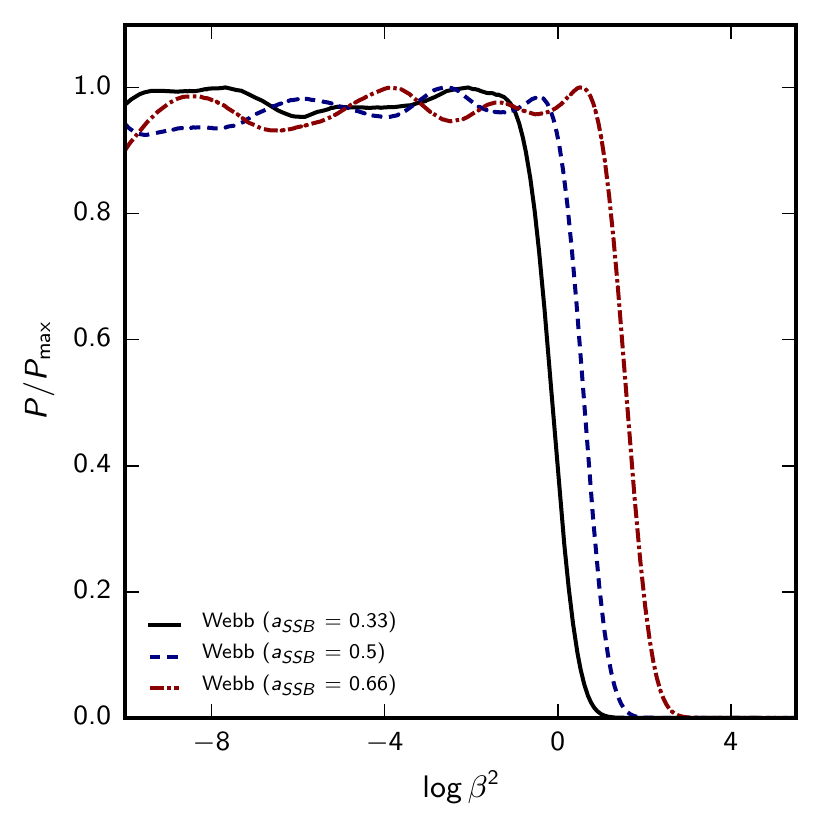}&
\includegraphics[width=0.5\columnwidth]{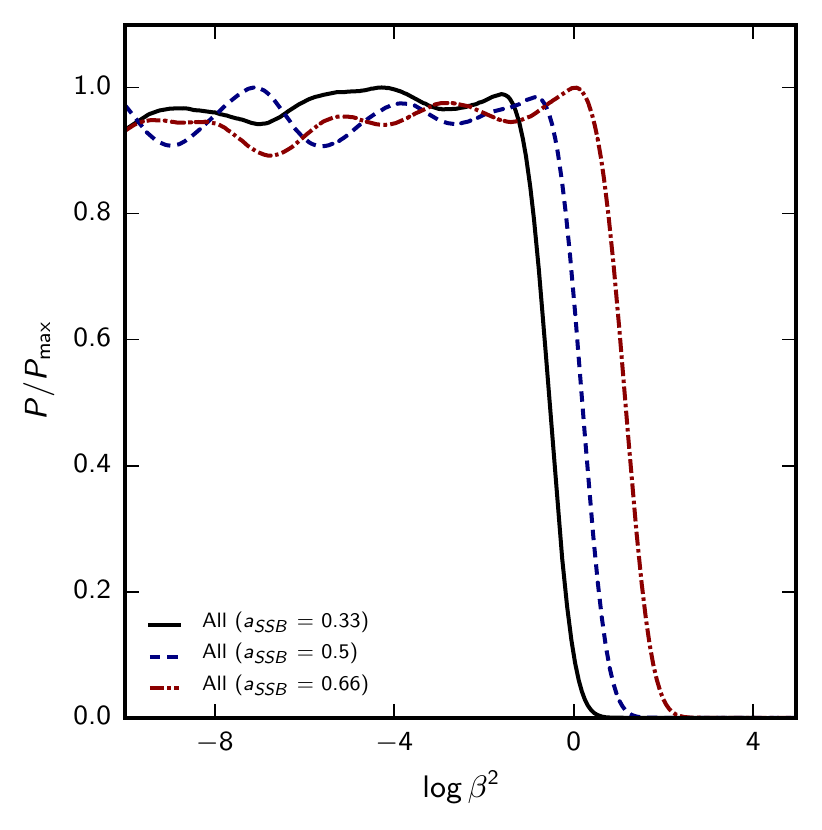}\\
\end{tabular}
\caption{{\it Top panel:} posterior distribution for the $a_{SSB}$ parameter, using the \webb\ dataset (blue solid lines) and the \all\ combination (red dashed lines). Here we have fixed $log(\beta^2)=1$ and $\lambda_{\phi 0}=1$. {\it Bottom panels:} posterior distribution for the $log \beta^2$ parameter with different values of $a_{SSB}$. On the left panel we use the \webb\ dataset, and on the right panel the \all\ combination. In all both cases we keep $\lambda_{\phi 0}=1$ fixed. }
\label{plt.beta1p}
\end{center}
\end{figure}

We also perform our analysis with only one free parameter, fixing $log \beta^2 = 1$ and $\lambda_{\phi0}=1$ Mpc/h, again on the grounds that they were used in the N-body simulations of \cite{marvin}, we find that the \webb\ dataset is not able to constrain $a_{SSB}$, while for the \all\ combination we find 
a $2-\sigma$ lower limit $a_{SSB}>0.43$, as displayed in the top panel of Figure \ref{plt.beta1p}. 

On the other hand, considering $log \beta^2$ as the only free parameter while fixing $\lambda_{\phi0}=1$ Mpc/h and different values of the epoch of symmetry breaking (specifically $a_{SSB} = 0.33, 0.5$ and $0.66$), the results shown in Figure \ref{plt.beta1p} and Table \ref{tab.res-1p} show that, as expected, if the symmetry breaks more recently a larger coupling value is allowed. Again the recent measurements improve the constraints from the archival measurements and, also in this case, we find that the results are consistent with a vanishing $\beta$.\\

\begin{table}
\begin{center}
\begin{tabular}{cccc}
\hline
{ }       & $a_{SSB}=0.33$ & $a_{SSB}=0.50$ & $a_{SSB}=0.66$ \\
\hline \hline
\webb\      & $< -0.5$       & $< 0.2$        & $< 1.2$ \\
\all\  & $< -0.9$       & $< -0.2$       & $< 0.7$ \\
\hline
\end{tabular}
\caption{Two-$\sigma$ constraints on the symmetron parameter $log \beta^2$ given by the \webb\ dataset and the \all\ combination, for different fixed values of $a_{SSB}$; $\lambda_{\phi0}=1$ Mpc/h was also used throughout.}
\label{tab.res-1p}
\end{center}
\end{table}

\section{Conclusions and outlook}\label{sec:conc}

In this work we have introduced a new methodology to accurately test models with spatial and/or environmental (local density dependent) variations of the fine-structure constant $\alpha$. These are based on the calculation of the angular power spectrum of these measurements, which are standard in other cosmological contexts. For concreteness we have also applied these tools to the case of $\alpha$ variations in symmetron models. We find that currently available data are not able to constrain the symmetron parameters $a_{SSB}$ and $\log{\beta^2}$ when they are both considered as free parameters. If instead the only free parameter is the strength of the coupling 
to gravity $\beta^2$, we find that the data do not show any deviations from the standard behavior and rather provide an upper limit for this coupling, which is
$\log{\beta^2}<-0.9$ in the most constraining case considered here.

Our results highlight the fact that a relatively small number of stringent measurements--- the recent dedicated measurements discussed in \cite{Dedicated}---lead to stronger constraints when they are combined with the larger dataset of earlier measurements. The current best constraints on the parameter $\beta$ come from pulsar timing constraints on Brans-Dicke type scalar tensor theories (of which symmetrons are an example), which correspond to $\beta\lesssim 10^{-2}$ \cite{symcosmo}. While our constraint is weaker, it comes from $\alpha$ measurements alone. Combining this with other cosmological datasets will lead to more stringent constraints; we leave this extended analysis for subsequent work.

Our results should be seen as a proof of concept, in the sense that they are limited by the uncertainties of the available $\alpha$ measurements. Future high-resolution ultra-stable spectrographs, in particular ESPRESSO \cite{ESPRESSO} (due for commissioning at the combined Cound\'e focus of the VLT in 2017) and ELT-HIRES \cite{HIRES} (for the European Extremely Large Telescope, whose first light is foreseen for 2024), both of which have these measurements as a key science and design driver, will lead to significantly more sensitive measurements, both in terms of statistical uncertainties and in terms of control over possible systematics. 
As discussed above, the use of more sensitive data and complete surveys will require a further step in the analysis, revising the assumptions made here for the calculation of both the theoretical and observational power spectra. On the other hand, these will enable more detailed studies, including a tomographic analysis (dividing the data into several different redshift bins) and, should variations be confirmed, model selection studies comparing various possible theoretical paradigms. A discussion of these possibilities is left for subsequent work.

\section*{Acknowledgments}

We are grateful to Luca Amendola, Micol Bolzonella and Adi Nusser for several discussions and useful suggestions during the various stages of this work.

This work was done in the context of project PTDC/FIS/111725/2009 (FCT, Portugal). CJM is also supported by an FCT Research Professorship, contract reference IF/00064/2012, funded by FCT/MCTES (Portugal) and POPH/FSE. MM is supported by the Foundation for Fundamental Research on Matter (FOM) and the Netherlands Organization for Scientific Research / Ministry of Science and Education (NWO/OCW). MM and CJM acknowledge additional support from the COST Action CA1511 (CANTATA), funded by COST (European Cooperation in Science and Technology).

\bibliographystyle{model1-num-names}
\bibliography{symmetron_v2}

\end{document}